# Piezo-resistive pressure sensor based on CVD-grown ZnO nanowires on Polyethylene Tetrathalate substrate

Manisha Kumari[1], Rakesh K. Prasad[1], Manish K. Singh[1], Parameswar K. Iyer[2], Dilip K. Singh[1*]

[1]*Department of Physics, Birla Institute of Technology Mesra, Ranchi-835215 (India)*

[2]*Department of Chemistry, Indian Institute of Technology Guwahati-781039 (India)*

*Email: dilipsinghnano1@gmail.com*

## Abstract

Recent developments in the domain of electronic materials and devices have attracted the interest of researchers toward flexible and printable electronic components like organic transistors, printable electrodes and sensors. Zinc Oxide (ZnO) nanowires (NWs) possess a number of excellent properties like high mobility, large exciton binding energy and the direct-band gap in addition to large piezoelectric coefficients. Here, we report on flexible piezo-resistive sensor based on Indium tin oxide (ITO)-coated Polyethylene tetrathalate (PET) substrate. The device shows sensitivity in terms of change in resistance from 100 to 2.4 K at an applied potential of 5V upon bending from flat to 95 degrees. The 1-D nanowire flexible device in its flat state shows saturated output current. We observed ten folds enhanced variation as compared to previous reports. Improved sensitivity was observed in our experiments due to fewer defects in CVD-grown NWs as compared to others where hydrothermally grown nanowires were used. The methodology of device fabrication reported here requires less time and enables efficient devices for the realization of flexible and wearable technology.

**Introduction**

Flexible and wearable electronics have received a lot of attention in the field of personal health [1] and environmental monitoring systems with advancing technology based on the internet of things. Due to the superior ductility, flexibility, durability and low weight [2] these devices are suitable for healthcare [3], robotic touch [4], intelligent clothing [5] and biomechanics [6]. The flexible pressure sensor can sense external stimuli and is a potential candidate for electronic skin (E-skin) [7] Such flexible sensors can be used to monitor personalized health parameters with improved accuracy and can be attached to a person's wrist to capture the pulse waveform making a large leap towards wearable biomedical devices [8]. All pressure sensors are based on the piezo-electric effect, capacitance variation and piezo-resistance variation [6,9]. Piezoresistive sensors normally convert applied pressure to the sensor into a resistance signal [10] Due to their appealing benefits like simple signal collection, easy manufacturing, affordable pricing and functionalities, pressure sensors hold immense potential for various applications [11]. In piezoresistive sensors, changes in contact resistance would occur either inside the materials of the active layer or between the conductive material and electrodes on the substrate. Therefore, their performance is influenced by two primary factors i.e., the base material (substrate) and the active components that are necessary for creating piezo-resistive sensors. A wide range of active materials has been used including carbon nanotubes [12], conductive polymers [13] and conductive nanowires [14]. A number of semiconducting materials has been used as piezo-resistive pressure sensors like GaN, CdS, ZnS and ZnO. Among these semiconducting materials ZnO is preferred over other materials because of low toxicity [15] and biodegradability [16]. ZnO holds exceptional semiconducting and piezoelectric capacity originating from its non-centrosymmetric crystal structure [17], polar crystal surfaces characteristics [18], presence of anharmonic phonons [19], Wide-bandgap, large exciton binding energy of 60 meV, piezo-electric coefficient [20]. Since 1-D ZnO NWs are utilized to make flexible electronics, they have a significant impact on soft wearable electronic systems. Due to its novel hierarchical structure and self-powered behavioral traits, including a high surface-to-volume ratio, high thermal and chemical stability, high electron mobility, high piezoelectric coefficient and high binding energy it is preferred as active material. Additionally, their low cost, biocompatibility, environmental friendliness and non-toxic qualities make them useful in piezotronics devices. [21].

There are numerous ways to synthesize ZnO nanowires namely hydrothermal growth [22-25], Sol-gel [26] and Chemical vapor deposition (CVD) [27]. In different synthesis techniques based on solution phase, the physical stability and low quantity of grown ZnO NWs are the major limitations. For example, the nanostructures produced using hydrothermal aqueous solutions may aggregate to form larger structures and precipitate at different rates. This technique requires use of variety of chemicals as precursors. To maximize the purity of the finished products, the majority of chemically precipitated nanoparticles need stabilizers and several post-processing steps. [28]. In contrast to the hydrothermal growth process, CVD-grown NWs produced at high temperature with high substrate homogeneity. They are highly stable, have superior crystalline quality, uniform size and shape and orientations of NWs can be controlled by adjusting growth parameters that can be controlled for relatively large mass production of 1D NWs.

In the present work, CVD grown ZnO NWs were used to fabricate self-powered flexible pressure sensor over an ITO-coated PET substrate. As fabricated device worked on the principle of piezoelectric effect of ZnO NWs and no external voltage source was needed for operation of the device. The functionality of the device is demonstrated in terms of change in resistance in the relaxed and bent positions of the device mounted on wrist. The device fabrication method reported is simple and utilizes spin-coating method to transfer NWs from rigid to the flexible substrate and is scalable for industrial production. The flexible pressure sensors reported would be highly useful in wearable health monitoring technologies and human-machine interfaces.

## Experimental section

## Growth of ZnO NWs

P-type silicon (Si) wafers (Si/SiO$_2$, 285 nm) were used as substrate. The wafer was cleaned under an ultrasonic bath and sequentially using Isopropyl alcohol (IPA), distilled water (DI) and ethanol solution. Post-cleaning 20 nm gold (Au) thin film was deposited using sputtering technique (Quarum Q150R ES) and was annealed at 600 °C for 15 minutes to yield Au-nanoparticles. ZnO nanowires were grown through vapor-liquid-solid (VLS) mechanism using the Chemical vapor deposition (CVD) technique using Techno OTF-1200X. The quartz boat was loaded with annealed Au/Si substrate, ZnO powder mixed with graphite powder in equal amounts in the center of a quartz tube of a thermal CVD reactor. It was subjected to 900°C to generate the Zn-vapour under a steady flow of argon gas (20 sccm) for 01 hour for the growth of NWs as shown in Figure 1(a).

## Device Fabrication

Fig.1 schematically shows the procedure of the device fabrication. After growth, the Si substrate with grown NWs was submerged in 200 µl of ethanol and sonicated for two hours to separate ZnO NWs from the substrate as shown in Figure 1(b). ITO-coated PET substrate (Sigma Aldrich, 60 Ω /sq., coated with 100 nm ITO) was taken as the flexible substrate. In this substrate, ITO acts as the bottom electrode. The PET substrate was sequentially cleaned using acetone, IPA and distilled water. Cleaned substrates were subjected to ozone treatment for 20 minutes at room temperature (NOVASCAN PSD Pro Series) for creating a hydrophilic surface [29]. To avoid damage to the substrate during spin coating, the cleaned glass slide was adhered to the flexible substrate using scotch tape and was placed over the spin coater. Spin-coating (Model WS-650MZ-23NPPB) was done for 30 seconds at 3000 rpm to deposit ZnO NWs on a flexible substrate as shown in Figure 1(c). As coated substrate was cured at 100 °C to remove the ethanol. The solution for dielectric layer was made using mixture of 3 ml of Chloro-Benzene (Sigma Aldrich) and 180 mg of PMMA (Sigma Aldrich) powder combined together and stirred for 6 hours using a magnetic stirrer to achieve a 60% homogenous solution of PMMA. Further, a thin layer of PMMA was coated using 200 µl of 60% homogenous solution of PMMA at 3000 rpm for 60 seconds. Post-coating, it was cured at 120 °C over a hot plate under an atmospheric environment to remove the solvent. An adhesive layer of 5 nm Titanium was deposited over the PMMA substrate using RF magnetron sputtering (M/S Advanced Process Technology, India). During sputtering the base chamber pressure was maintained at $1 \times 10^{-5}$ mbar and pressure was kept constant throughout the deposition by varying the sputtering gas, which was high-purity argon (99.99 %) for two hours at a constant RF power of 80 W. The top gold electrode was deposited over the adhesive thin titanium layer by thermal evaporation (Model No. HHV AUTO 500AW) at $6 \times 10^{-6}$ mbar chamber pressure using thermal evaporator. The device was then cured on a hot plate at 50 °C for 20 minutes under yellow light. The different layers of the fabricated flexible device is schematically shown in Figure 1(d). It consists of ITO-coated PET substrate which is flexible in nature. ZnO NWs (active material) are spin-coated on the top of the bottom electrode. Then, the dielectric layer of PMMA is spin-coated on top of active material. PMMA was used to provide a compact piezo layer with minimal porosity, low roughness, uniform thickness and good piezoelectric characteristics that could be successfully utilized in a functioning piezo-electric device. Also, it fills the spaces between the ZnO nanowires and takes the place of the Schottky barrier [30].

After that titanium is sputtered on top of dielectric material which acts as an adhesive layer between the top electrode and the dielectric layer. At last, gold is deposited as the top electrode. Due to the ductile nature, gold electrode doen¢t undergo crack while bending. Finally fabricated device is shown in Figure 1(e).

**Characterization**

The surface morphology and alignment of the grown NWs were characterized using Field Emission Scanning electron microscopy (FESEM), Model No: Zeiss Sigma 300. The elemental analysis of NWs is performed using Energy Dispersive X-ray (EDX) spectrometer (EMS-850) coupled with FESEM. The structural properties of the grown samples were characterized using Smart Lab, Rigaku, Japan X-ray diffraction diffractometry (XRD) at a scan rate of 5º/min. The optical absorption spectra of the NWs were measured using a UV-Visible Spectrometer (PerkinElmer UV/VIS/NIR Spectrometer Lambda 950) within the spectral range of 200-900 nm. The fluorescence of NWs was characterized by photoluminescence spectrometer with $\lambda_{ex}$ = 360 nm from He-Cd laser and using Flurolog-3, Horiba at room temperature**.** The optical transmittance spectra of the flexible substrate and device were measured using a UV-VIS spectrophotometer (Model- UV 1900i, Shimadzu) at room temperature. For electrical measurements, Semi-Conductor Parameter Analyzer (Model no. Keithley 4200 SCS Pulse Parametric Analyzer) probe station (model: Ever being C-6 Probe station) was used with a tungsten probe with a diameter size of 10 microns. The piezoelectric coefficients $d_{33}$ of NWs were measured by the direct piezoelectric method through the YE2730A $d_{33}$ meter.  We measured the output current from ZnO NWs-based flexible devices under vertical compressive strains under different bending angles using a custom-made Piezoelectric setup. All of the piezoelectric output currents generated from the device were measured using a source meter Keithley 4200 semiconductor characterization unit.

**Device performance set-up**

The constructed sensor was fixed in the mid area of a mechanical vibrator (PASCO Model SF 9324) that was coupled to a power amplifier (POSCO Scientific Model CI-6552A). An arbitrary function generator (Tektronix AFG1062) was used to determine the vibrator's frequency and amplitude. This function generator produced amplified electrical signals with a predetermined amplitude, frequency and shape. The upper and bottom electrodes of the manufactured sensor were linked to a two-channel digital storage oscilloscope (Tektronix TDS1001B), which was used to gauge the output voltage.

**Result and discussion**

The structural morphology of CVD grown ZnO NWs is shown in Figure 2 with the help of FESEM images and EDX graph. The top view of ZnO NWs is shown in Figure 2(a) which indicates that the grown NWs are highly dense. The side view in Figure 2(b) indicates that ZnO-NW arrays were vertically grown and were randomly aligned. NWs were of length ~ 2.6 μm as estimated from FESEM using ImageJ software (inset of Figure 2(b)). These NWs were randomly grown due to significant lattice mismatch between the ZnO NWs and the entire Si substrate [31]. Figure 2(c) is a magnified image of ZnO NWs which shows that all NWs have uniform thickness. The average diameter (d ~108 nm) of NWs was estimated through ImageJ software. EDX spectra as shown in Figure 2 (d) indicate the presence of zinc (Zn) and oxygen (O) in NWs. It indicates the absence of any impurities in our CVD-grown ZnO NWs. Since the NWs formation is controlled by the vapor-liquid-solid (VLS) mechanism as described by Wagner and Ellis in 1964 for the formation of silicon whiskers [32]. Catalyst-assisted growth of ZnO NWs through VLS mechanism takes place through different stages. First stage involves aggregation of Au small islands layers on post-annealed silicon substrates coated with gold (Au). It further develops to create Au nanodroplets (NDs). The second stage involves the carbothermal reduction of ZnO to produce the Zn vapor phase, which is then transported through Ar gas and reacts with the Au NDs to produce an Au-Zn alloy. These nanoalloy droplets continuously absorbs the Zn and $O_2$ vapor at growth temperature and interacts within the droplet until ZnO attains supersaturation. At this point, alloy droplets and substrate forms interface with precipitated ZnO nanocrystals. As a result, ZnO NWs were produced with Au nanoparticles present on the tip of these NWs. It has been observed that the catalyst size and position govern the diameter and position of the nanowire growth [33]. High-temperature growth results in high vapors with increased kinetic energy that diffuses more rapidly to create and improve nanowire quality suitable for nano-electronic applications.

The structural and optical characterization of figure 3(a) shows the X-ray diffraction (XRD) pattern of CVD-grown ZnO NWs at a scan speed of 5 degree/min. It indicates peaks at $31.74°$, $34.52°$ and $36.27°$ corresponding to (100), (002) and (101) lattice planes respectively. Peak corresponding to (101) plane was observed to be most dominant indicating formation of ZnO wurtzite structures. Sharp XRD line profiles indicats highly crystalline nature of as-synthesized ZnO NWs. The observed peaks correspond to the wurtzite hexagonal phase with lattice constants of a = b = 0.3288 nm and c = 5.2054 nm and belong to space group: P63mc (JCPDS card number 01-089-0510 for ZnO) [34]. The Bragg reflection of silicon substrate occurs at peak position $32.90°$ corresponding to lattice plane (200). Figure 3(b) shows Raman spectra of ZnO NWs with laser excitation of 532 nm. Since the ZnO crystal is wurtzite in structure belonging to $C_6^4V$ space group, the possible optical phonon modes at the zone center are:

$$_{opt} = 1A_1 + 2B_1 + 1E_1 + 2E_2$$

Where $A_1$, $E_1$ and $E_2$ are characteristics of Raman active modes, whereas $B_1$ is an indication of a Raman inert mode. As the polar modes $A_1$ and $E_1$ are divided into the transverse optic (TO) and longitudinal optic (LO) phonons. The $E_2$ mode produces two peaks for the

$E_2^{(high)}$ and $E_2$ (low) mode phonons and is associated with the non-polar phonon vibration [35]. We observed peaks at 99 cm$^{-1}$, 437 cm$^{-1}$, 521 cm$^{-1}$ and 583 cm$^{-1}$ in which first two peaks are associated with $E_2$ mode namely $E_2^{low}$ mode and $E_2^{high}$ mode respectively. $E_2^{low}$ mode corresponding to vibration mode in Zn sub-lattice [36] and $E_2^{high}$ mode represents the characteristics of wurtzite phase [37]. The intense peak at 521 cm$^{-1}$ arises from the silicon substrate [38]. The $E_1$(low) mode observed at 583 cm$^{-1}$ due to second order Raman scattering. Observed Raman spectra is in agreement with XRD study, which demonstrates the wurtzite structure of ZnO NWs. The UV-Vis absorption spectra for ZnO-NWs are shown in Figure 3(c) which is ultrasonically dispersed in ethanol. With an absorption of around 60% of the incident UV light, the ZnO-NWs solution exhibits the typical UV sharp absorption peak centered at 374.7 nm (3.30 eV), which is close to the energy gap of bulk ZnO [39]. The peak observed at 260 nm is far below the energy gap of ZnO. It primarily results from formation of nanowire of ZnO. This indicates potential application of formed ZnO-NWs as UV absorber and charge carrier-producing material. The PL emission spectra of ZnO NWs at room temperature are shown in Figure 3(d). The photoluminescence (PL) characteristics of ZnO NWs shows a clear relationship between the level of oxygen shortage in the crystalline structure of the material and the intensity of the visible emission of ZnO. This means that PL measurements can be used to quantify the crystalline defects present in the ZnO NWs crystalline structure. The emission bands in the UV and visible regions of the spectra have asymmetrical line shapes. The spectra were fitted with the Gaussian line shape to find the peak position. Bound excitonic transitions in ZnO NWs are linked to the peak at 379 nm (3.27 eV), which is less intense and narrowly known as the near band-edge (NBE) emission. Typically, the NBE emission acts as an indicator of the sample's intrinsic crystalline quality. The FWHM of this NBE emission is 12.4 nm, which is also compatible with the Raman data. This NBE emission was observed to confirm the quality of grown NWs. Additionally, a visible broad and intense emission band in the range

400 and 700 nm was observed. This emission band is typically attributed to deep-level defects including vacancies and interstitials of zinc and oxygen [40]. Surface defects like oxygen vacancies (VO) results into the green band at ~ 536 nm [41].

Figure 4(a) represents the FESEM picture of ZnO nanowire arrays that were separated from the Si substrate before being smeared on an ITO-coated PET substrate. The NWs distributed evenly on the ITO/PET substrate overlaps with each other and acts as welded junction. The crystallographic orientations of the NWs in the subsequent device are aligned in the same direction due to the overwhelmed nanowire arrays having the same growth direction. As a result, all of the ZnO nanowires contribute positively to the macroscopic potential, which is determined by the polarities of the induced piezo-potentials of these nanowires oriented in the same direction. By using the direct piezoelectric method, the piezoelectric coefficients $d_{33}$ of ZnO NWs were measured as 7pC/N. Figure 4(b) shows the optical transmittance spectra of the ITO/PET substrate, which is taken to the reference as glass substrate. The flexible conducting substrate is transparent in nature (inset 4b(i)). However, for spin-coated ZnO on ITO/PET substrate, the transmittance decreases by 0.91% due to ZnO NWs smeared on conducting substrate. The fabricated device (inset 4b(ii)) shows decrease of transmittance to 30 % in the visible region due to low transparency (semi-transparent). Figure 4(c,d) shows the wearability test of the flexible device performed by the Keithley 2400 multi-meter in which the device is fixed on the wrist of a hand. Figure 4(c) shows the optical image of releasing mode (flat) of the flexible device. During this mode, the device's resistance is 0.1 KΩ . Figure 4(d) shows an optical image of the convex bending of the device. During bending mode, the device shows resistance of 2.4 kΩ . This is due to the compressive force applied on the device causing tensile strain on NWs. When the external force is applied to the device, bound charges will be released, generating piezoelectric voltage between the top and the bottom electrode layers [42]. Figure 4(e) shows the open-circuit voltage with respect to time. The voltage versus time plot of the flexible device shows peaks on the positive voltage value during the bending stage at zero bias voltage and in the negative voltage value during releasing stage. The ZnO-based flexible device's mechanism for producing voltage is provided by the mechanical stress that is created by vibrational motion between NWs. In the ZnO NWs film of the wurtzite structure, which gives the $Zn^{+2}$ cations a displacement with respect to the $O^{-2}$ anion. The two electrodes of the device collect these electrical charges.

The bending angle can be determined by the bending arc (diameter) made by the flexible device as shown in Figure 5(a). Since bending results in the entire electrode network being elongated, resulting into increase in resistance by decreasing the bending angle as shown in Figure 5(a). We measured piezo-resistive output current without external bias using a custom-made pressure exerting system. The results show that the fabricated device exhibits piezoelectric characteristics in response to a uniform external force. According to our results, the piezo-resistive characteristics of this sensor depend upon the nature of ZnO NWs, since the electric dipoles in the NWs oscillate arbitrarily within a degree of their respective aligning axes while maintaining in a macro equilibrium condition, there is initially no piezoresistive output when there is no impact or vibration on the sensor. A potential difference originated between the two electrodes as the pressure sensor was crushed vertically by a mechanical impact and the flow of electrons is generated between the two electrodes. Using the above-mentioned experimental setup, the fabricated sensor is systematically characterized. First, the energy harvesting capability of flexible devices is evaluated for a range of compressive pressure amplitudes. The voltage generation for piezoresistive sensors are directly correlated with the applied stress as evident from change in current with increasing bending angle. In accordance with the rise in compressive pressure amplitude from 45 kPa to 61 kPa, as measured across a resistive load of 120 Má , the open circuit voltage ($V_{oc}$) amplitude increases from 1.5V to 7.5V as shown in Figure 5(d).

**Conclusion**

We report a simple and low-cost fabrication approach for a ZnO NWs-based device with potential applications in flexible piezo-resistive pressure sensors. The ZnO NWs were created using a CVD technique to generate high-quality length ~ 2.6 μm, vertically aligned ZnO nanowires with diameters of 108 nm, which were then spin-coated on an ITO/PET substrate. High temperature growth results into high vapor with increased kinetic energy that diffuse more rapidly to create and improve nanowire quality suitable for nano-electronic applications. The XRD and RAMAN results confirms the hexagonal wurtzite crystal structure of grown ZnO NWs. Furthermore, UV-Visible absorption peak around 3.30 eV confirms the NWs acts as UV rays absorbing material. The PL spectra confirmed the ZnO NWs' decent crystal quality due to narrow near band edge emission around 12.4 nm. Apart from hydrothermal and photolithography processes, spin-coating techniques that do not affect the flexible substrate are employed to transfer ZnO NWs in flexible devices. The fabricated device is semi-transparent in nature since its transmittance spectra is 30% in visible region. These devices can reduce their fabrication costs by maintaining adequate electromechanical performance to supply electrical energy due to NWs have piezoelectric coefficient $d_{33}$ as 7Pc/N. Finally, we showed that our ZnO NWs flexible device fabrication technique could be scaled up for large film applications and applied to new 1D flexible systems which shows sensitivity around 0.18V/kPa. Zinc oxide (ZnO) nanowires (NWs) are an excellent candidate for mechanical sensors and energy harvesters used in piezotronics and piezophototronics devices.

**ACKNOWLEDGMENT**

Dilip K. Singh thanks UGC-DAE CSR Indore (CRS/2021-22/01/358), NM-ICPS ISI Kolkata and DST, Government of India (CRG/2021/002179; CRG/2021/003705) for financial support. We are also thankful to the CIF, BIT Mesra and INUP of IIT Guwahati (Project No. INUP project code MeitY No.5(1)/2021-NANO funded by MeitY) for their support in terms of facilities and equipment.

**Figure captions:**

Figure 1: Schematic of the experimental procedure adopted for device fabrication. (a) ZnO NWs grown on Si substrate. (b) Schematic of the ultra-sonication process to detach NWs from the rigid substrate. (c) Spin coating of NWs on the flexible substrate. (d) Schematic diagram of different layers of fabricated flexible device. (e) Photograph of the final fabricated device.

Figure 2: FESEM images of the ZnO NWs grown on the Silicon substrate by the CVD process (a) Top-view (b) Side view of CVD-grown ZnO NWs (c) magnified image of ZnO NWs (d) EDAX of CVD-grown ZnO NWs.

Figure 3: (a) XRD image of as-grown ZnO NWs. (b) Raman spectra of ZnO NWs. (c) Absorbance Spectra of ZnO NWs through 360 nm laser. (d) Photoluminescence spectra of grown ZnO NWs using $\lambda_{ex}$= 360 nm laser.

Figure 4: (a) FESEM image of detached ZnO NWs. (b) Transmittance spectra of ITO/PET & ZnO NWs coated ITO/PET, fabricated device. (c) Flat working mode of the device. (d) Bending working mode of the device. (e) Voltage versus Time plot of the flexible device at zero bias voltage (open-source voltage) (f) Principle of piezoelectric output in releasing mode of the flexible device (g) Principle of piezoelectric output in bending mode of the flexible device.

Figure 5: (a) I-V of flexible Device. Inset shows bending set-up for I-V measurement (b) Variation of bending angle versus current. Inset shows schematic of the bending of the device. (c) Open circuit output voltage with respect to the applied compressive pressure (d) Summarized variation at 5 Hz frequency of peak open circuit voltage.

**Figures:**

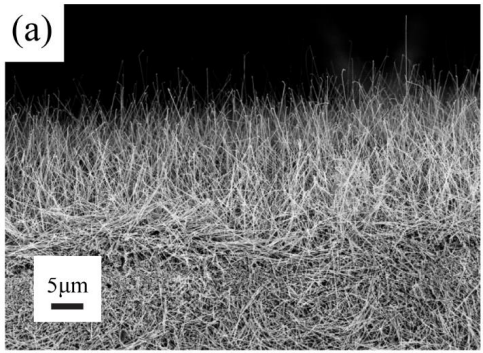

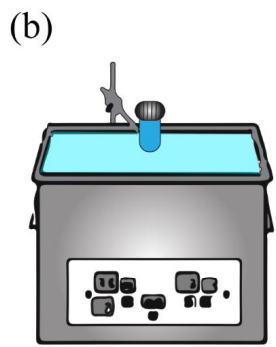

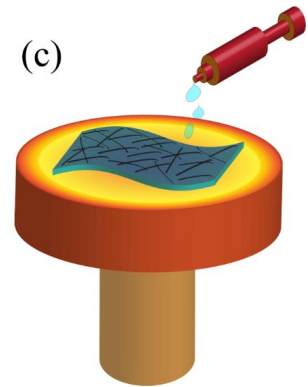

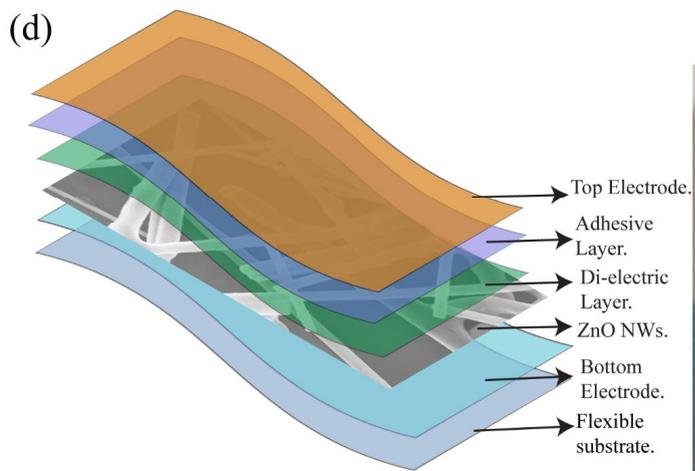

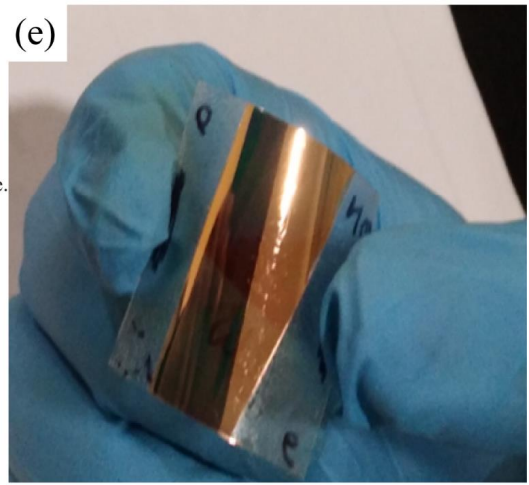

**Figure 1** Singh et al.

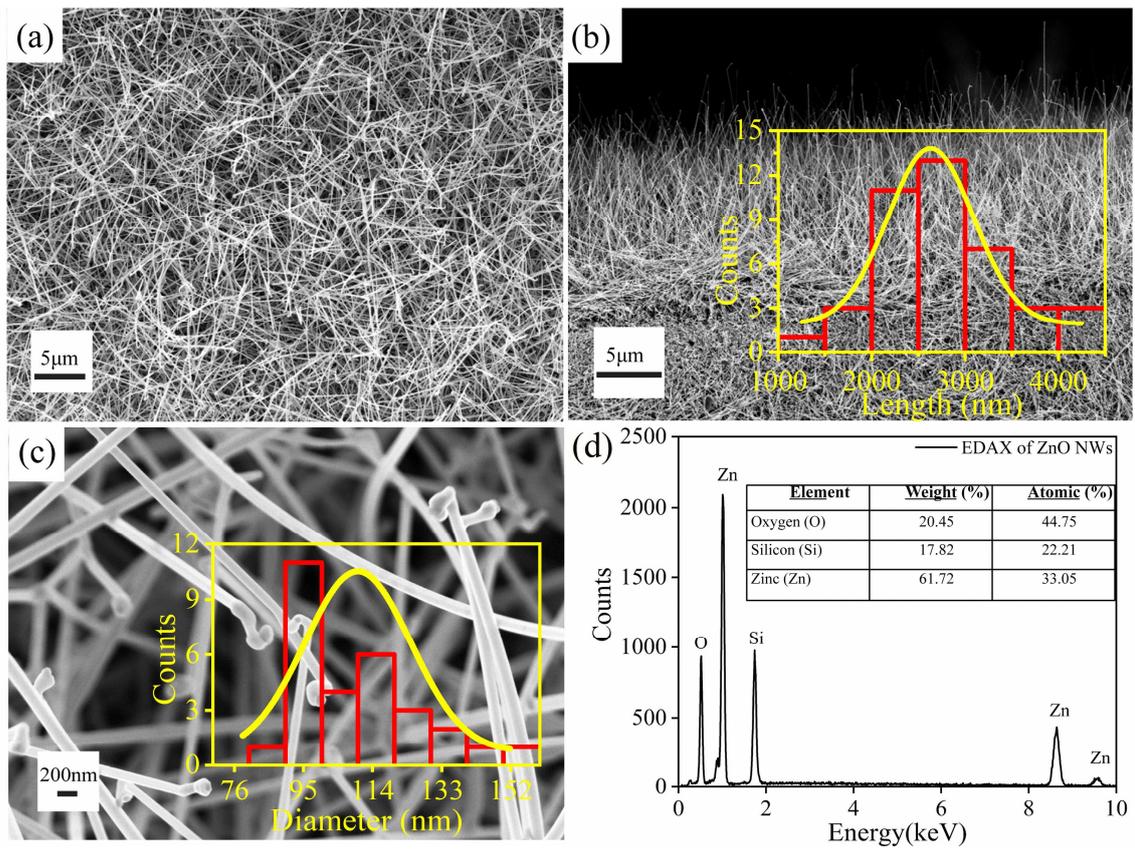

**Figure 2** Singh et al.

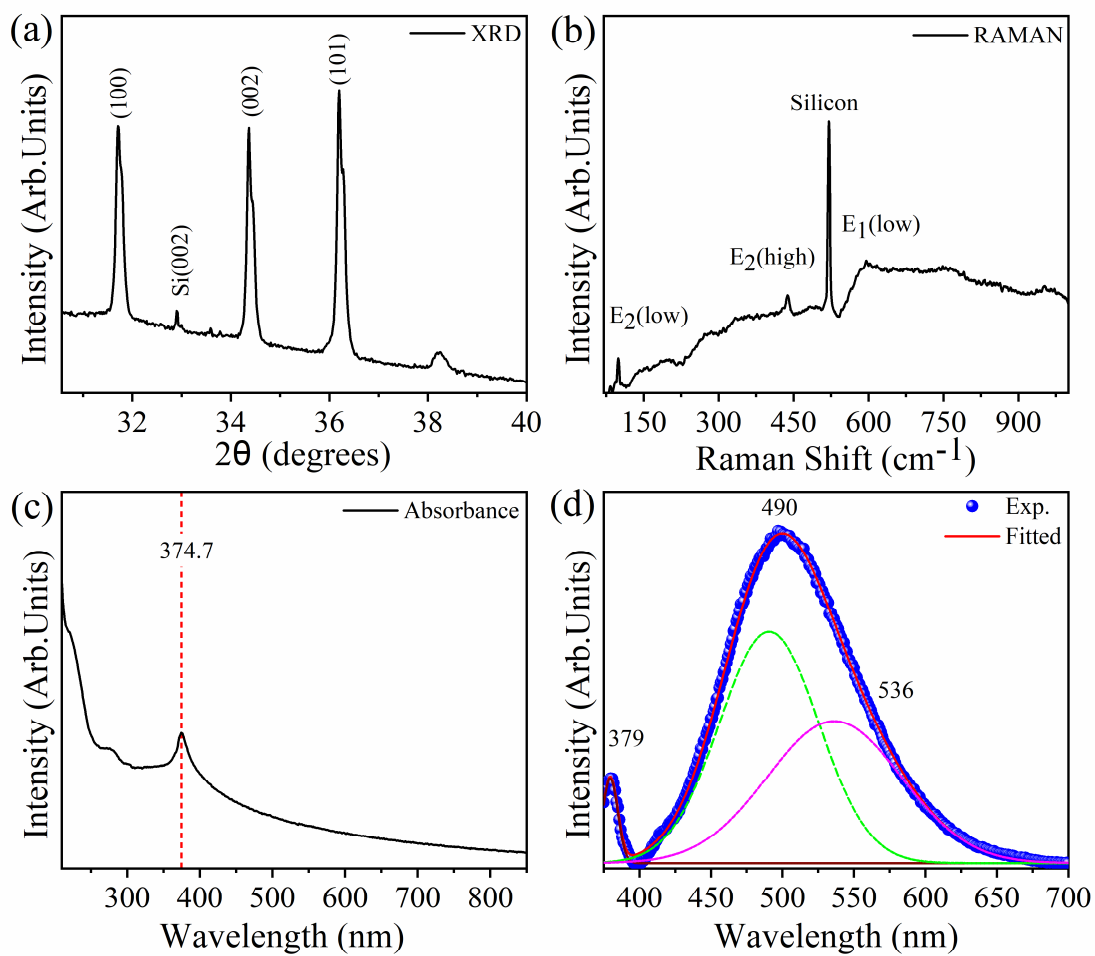

**Figure 3** Singh et al.

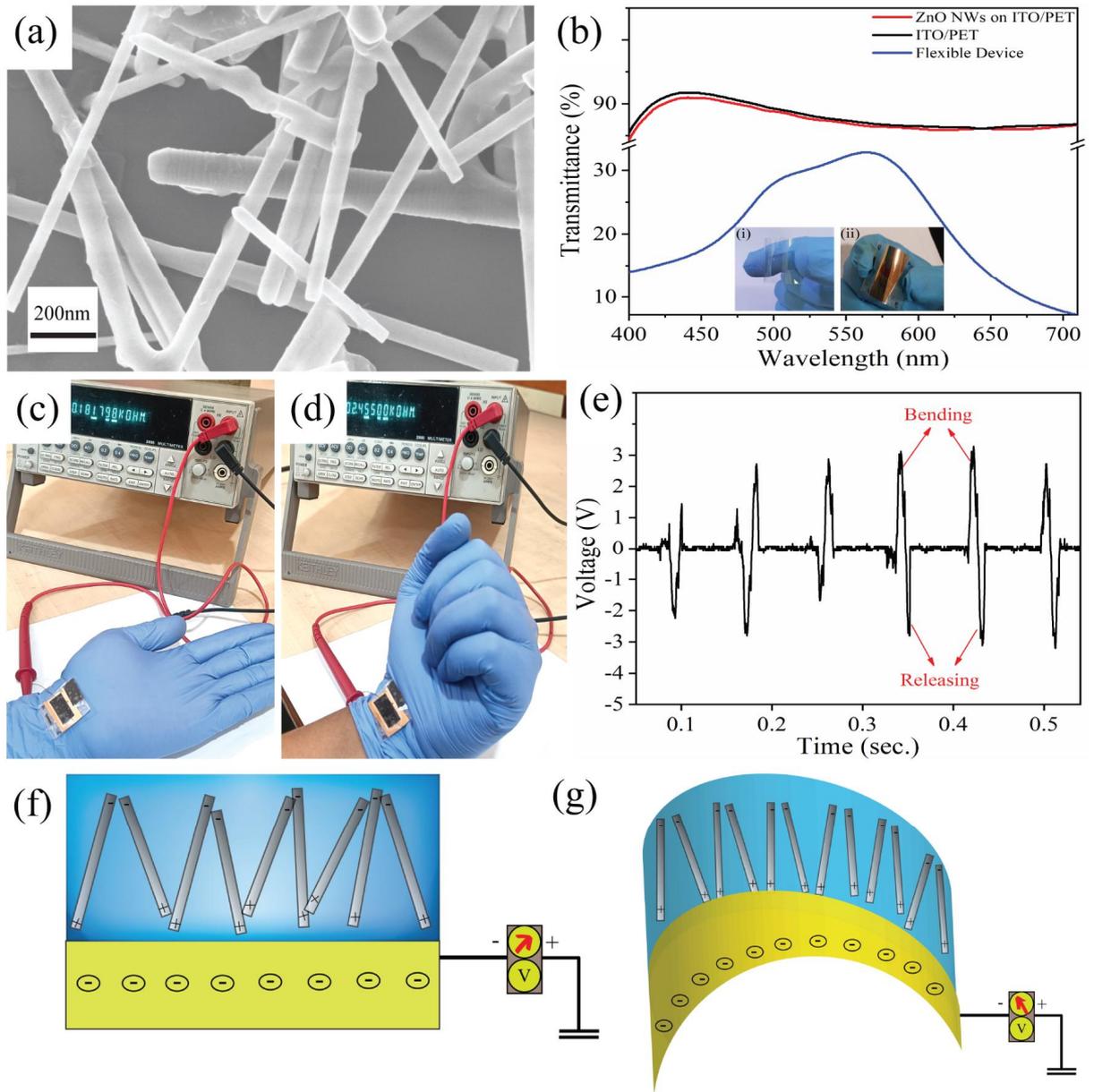

**Figure 4** Singh et al.

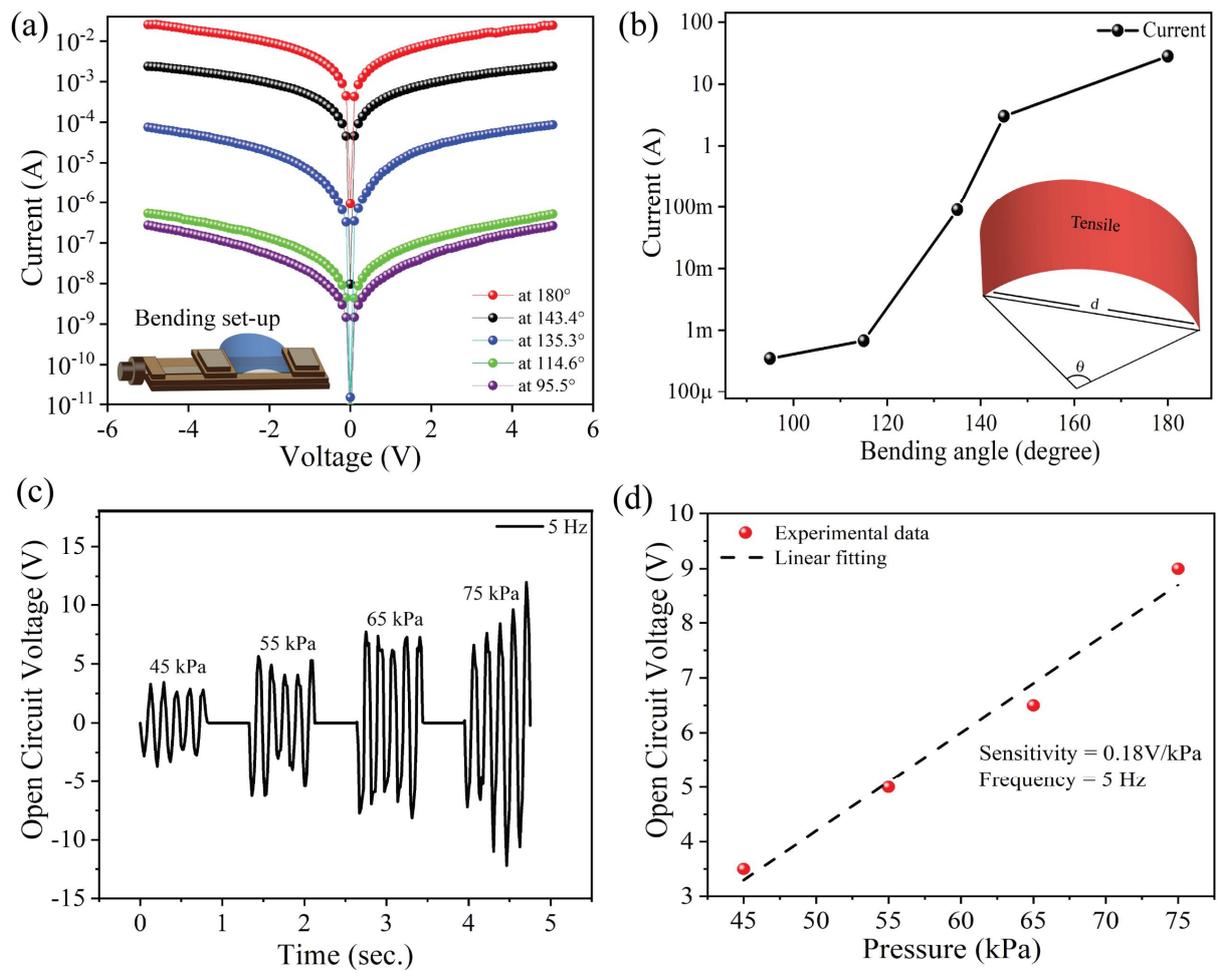

**Figure 5** Singh et al.